\documentclass[preprint,12pt,authoryear]{elsarticle}

\usepackage{amssymb}
\usepackage{array}
\usepackage{amsmath}
\begin{document}

\begin{frontmatter}

\title{Capabilities of object-oriented programming for the construction of quantum-kinetic BBGKY equations of high orders} 

\author{Tarasevich E. A.} 

\affiliation{organization={P.N. Lebedev Physical Institute RAS, Branch in Troitsk},
            addressline={Fizicheskaya Ulitsa, 11}, 
            city={Moscow, Troitsk},
            postcode={108840}, 
            country={Russia}}

\affiliation{organization={ E.V. Shpol’skii Theoretical Physics Chair, Moscow Pedagogical State University (MPGU)},
            addressline={Malaya Pirogovskaya Ulitsa, 1/1}, 
            city={Moscow},
            postcode={119435}, 
            country={Russia}}

\author{Gladush M. G.} 

\affiliation{organization={P.N. Lebedev Physical Institute RAS, Branch in Troitsk},
            addressline={Fizicheskaya Ulitsa, 11}, 
            city={Moscow, Troitsk},
            postcode={108840}, 
            country={Russia}}
            
\affiliation{organization={ E.V. Shpol’skii Theoretical Physics Chair, Moscow Pedagogical State University (MPGU)},
            addressline={Malaya Pirogovskaya Ulitsa, 1/1}, 
            city={Moscow},
            postcode={119435}, 
            country={Russia}}
            
\begin{abstract}
Theoretical methods based on the density matrix are powerful tools to describe open quantum systems. However, such methods are complicated and intricate to be used analytically. Here we present an object-oriented framework for constructing the equation of motion of the correlation matrix at a given order in the quantum chain of BBGKY hierarchy used to describe the interaction of many-particle systems. The algorithm of machine derivation of equations includes the implementation of the principles of quantum mechanics and operator algebra. It is based on the description and use of classes in the Python programming environment. Class objects correspond to the elements of the equations that are derived: density matrix, correlation matrix, energy operators, commutator and several operators indexing systems. The program contains a special class that allows one to define a statistical ensemble with an infinite number of subsystems. For all classes, methods implementing the actions of the operator algebra are specified. The number of subsystems of the statistical ensemble for the physical problem and the types of subsystems between which pairwise interactions are possible are specified as an input data. It is shown that this framework allows one to derive the equations of motion of the fourth-order correlation matrix in less than a minute.
\end{abstract}





\end{frontmatter}


\section{Introduction}
Nowadays numerous studies of microworld are focusing on the interaction of radiation with matter, which is a problem of interaction of many particles. This problem, both in the classical and quantum-mechanical limits, can be considered in terms of a many-particle distribution function or quantum distribution function – a density matrix \citep{blum2012density}. The property of reduction of the distribution function and its quantum-mechanical analogue allows one to solve the problem of interaction of many particles by considering only a limited number of subsystems of the complete system. This approach consists in derivation of an equation for the reduced density matrices of one and several subsystems. In this case, the influence of the rest of the system is automatically taken into account through generalized terms – self-consistent fields and collision integrals. Calculating the latter is a separate theoretical problem which is solved by using physical assumptions and models. One of the standard methods of mathematical physics in this area is the construction of chains of Bogolyubov-Born-Green-Kirkwood-Yvon (BBGKY) equations, also known as the Bogolyubov chains methods. The BBGKY method is a complete and consistent approach to the study of many-particle systems and allows one to study the properties of individual particles under the influence of a self-consistent field generated by the entire system. For example, this approach was used to study the effect of a local field on the radiative relaxation rate of single quantum emitters in a dielectric medium \citep{kuznetsov2011_JETP, kuznetsov2011_TMP, MG2011_EPJD}. Recently, this method was applied to study the phonon dynamics in quantum emitters \cite{pandey2024}. However, the analytical derivation of high-order correlation functions takes a significant amount of time. At the same time, direct derivation of equations without making additional physical assumptions can be algorithmized. Thus, the aim of this work is to demonstrate a program code capable of deriving correlation matrices of a given order in a chain of BBGKY equations. The implementation of this idea is based on the object-oriented programming (OOP) architecture in the Python programming environment. The Python programming language is equipped with extensive libraries and is easy to read and use. OOP is suitable for large, complex and actively updated or supported programs. Elements of analytical output (density matrices, correlation matrices, commutators, and others) can be represented as classes with corresponding attributes, while methods of individual classes can perform actions of operator algebra.

\section{Density matrices to describe the evolution of quantum systems}

The density matrix method is a universal approach for describing both open and closed systems. There are several ways to treat density matrix. One of the most powerful is construction of BBGKY hierarchy. In this section this will be described on the example of light-matter interaction. Consider a system consisting of an ensemble of quantum emitters $\{ a \}$ and modes of a quantized electromagnetic field $\{ f \}$. It should be added that this approach is not limited to these sets of particles, but can be extended to larger types of particles, such as other types of emitters, phonon modes, etc. Consider the most general case, when the interaction between emitters occurs through electromagnetic field modes. Then the Hamiltonian of this system takes the form:
\begin{equation}
\hat{H} = \sum_a \hat{H}_a + \sum_f \hat{H}_f + \sum_a \sum_f \hat{V}_{af}, 
\end{equation}
where $\hat{H}_{a}$ are energies of free emitters, $\hat{H}_f$ - energies of free modes of electro-magnetic fields, $\hat{V}_{af}$ - interaction between the $a^{th}$ emitter and $f^{th}$ mode. The density matrix of all particles $\rho = \rho_{\{a\} \{f\}}$ completely describes this system, and its evolution obeys the von Neumann equation:
\begin{equation} \label{vN_sch}
i\hbar \frac{d}{dt} \rho = \left[ \hat{H}, \rho \right],
\end{equation}
with initial condition $\rho(t_0)=\rho_0$.

\begin{figure}[t]
\centering
\includegraphics[scale=0.2]{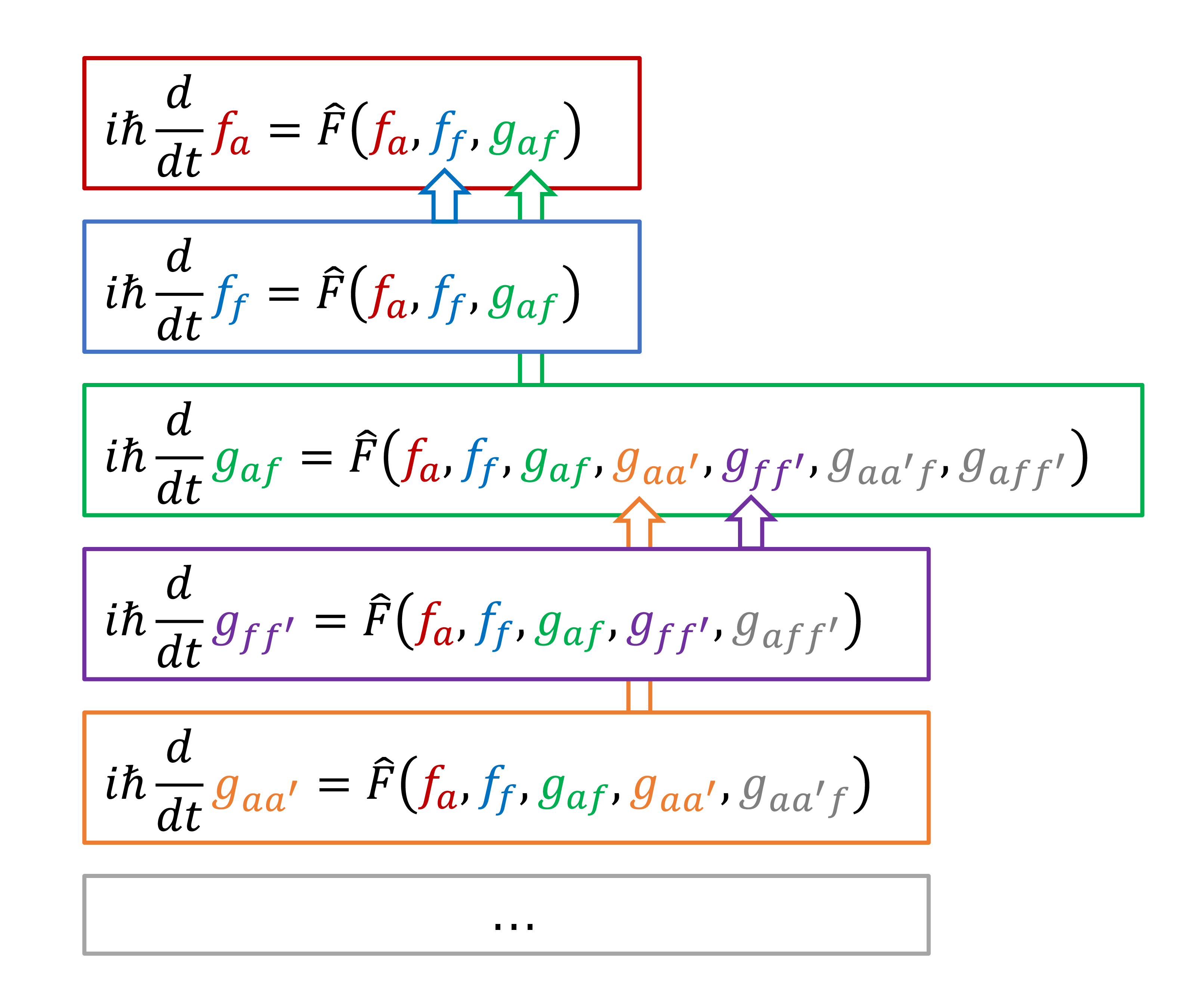}
\caption{Schematic diagram of further use of the BBGKY method for deriving the control equation.}\label{fig_1}
\end{figure}

The general density matrix is normalized to unity: $Tr_{\{a\}\{f\}} \rho=1$. Here and below $Tr_{\{s\}}$ denotes the operation of taking a trace over a set of particles $\{s\}$. Reduced density matrices $f_{\{s\}}$ describing one particle or a subsystem of particles ($n$-particles) can be found from the general density matrix $\rho$ through the operation of taking a trace over all indices that do not belong to the desired density matrix:
\begin{equation}
f_{\{a'\}\{f'\}}=Tr_{\{a''\}\{f''\}} \rho.
\end{equation}
Here the sets $\{a'\}$, $\{a''\}$,$\{f'\}$ and $\{f''\}$ together form the complete set of all emitters and modes of the quantized electromagnetic field in the system. In this case, the normalization condition for the reduced functions takes the next form:
\begin{equation}
Tr_{\{a'\}\{f'\}}f_{\{a'\}\{f'\}}=1.
\end{equation}
For reduced density matrices, the equations of motions are obtained by taking the trace of the von Neumann equation and can be written as the following recurrence relation:
\begin{eqnarray}
i\hbar \frac{d}{dt} f_{\{a'\}\{f'\}} - \left[ \sum_{a'} \hat{H}_{a'} + \sum_{f'} \hat{H}_{f'} + \sum_{a', f'} \hat{V}_{a'f'}, f_{\{a'\}\{f'\}} \right] = \\ \nonumber
=\sum_{a'', f'} Tr_{a''} \left[ \hat{V}_{a''f'}, f_{a'' {\{a'\}\{f'\}}} \right] + \sum_{a', f''} Tr_{f''} \left[ \hat{V}_{a'f''}, f_{{\{a'\}\{f'\}}f''} \right].
\end{eqnarray}
Here the left-hand side completely coincides with the von Neumann equation written for the reduced density matrix with the corresponding Hamiltonian. The right-hand side describes the connection of the particles under consideration with the rest of the particles in the system through all possible pairwise interactions. Next the specific examples of dynamical equation of reduced density matrices starting with a minimum number of particles are given:
\begin{equation}
i\hbar \frac{d}{dt} f_{a} - \left[  \hat{H}_{f},f_a  \right] = \sum_{f} Tr_{f} \left[ \hat{V}_{af}, f_{af} \right],
\end{equation}
\begin{equation}
i\hbar \frac{d}{dt} f_{f} - \left[  \hat{H}_{f},f_f  \right] = \sum_{a} Tr_{a} \left[ \hat{V}_{af}, f_{af} \right],
\end{equation}
\begin{eqnarray}
i\hbar \frac{d}{dt} f_{af} - \left[ \hat{H}_{a} + \hat{H}_{f} +  \hat{V}_{af}, f_{af} \right] = \\ \nonumber =\sum_{a'\neq a} Tr_{a'} \left[ \hat{V}_{a'f}, f_{aa'f} \right] + \sum_{f' \neq f} Tr_{f'} \left[ \hat{V}_{af'}, f_{aff'} \right],
\end{eqnarray}
\begin{equation}
i\hbar \frac{d}{dt} f_{aa'} - \left[ \hat{H}_{a} + \hat{H}_{a'}, f_{aa'} \right] =  \sum_{f} Tr_{f} \left[ \hat{V}_{af}+\hat{V}_{a'f}, f_{aa'f} \right],
\end{equation}
\begin{equation}
i\hbar \frac{d}{dt} f_{ff'} - \left[ \hat{H}_{f} + \hat{H}_{f'}, f_{ff'} \right] =  \sum_{a} Tr_{a} \left[ \hat{V}_{af}+\hat{V}_{af'}, f_{aff'} \right].
\end{equation}
This system of equations represents a chain of equations, where the evolution of each equation of reduced $n$-particle density matrix depends on the reduced $(n+1)$-particle density matrix. For further work with a chain of linked equations it is convenient to use the principle of weakening correlations and to represent multiparticle matrices in the form of a superposition of single-particle density matrices and correlation operators $g$ of $n$-particles:
\begin{equation} \label{cluster_dec}
f_{af} = f_a f_f + g_{af}, \;\;\;\; f_{aa'f} = f_a f_{a'} f_f + f_a g_{a'f} + f_{a'} g_{af} + f_f g_{aa'} + g_{aa'f}.
\end{equation}
The terms which contain products of single-particle density matrices  are corresponding to the case of weak correlation between particles. Multiparticle correlation operators $g$ correspond to the limit of strongly correlated particles in the result of their interaction. For subsystems of three or more particles, there are intermediate states when part of the system is strongly correlated while the other part of the system is weakly correlated with this part. In this cluster decomposition, the following conditions are imposed on the correlation matrix:
\begin{equation} \label{corr_mat_constrain}
Tr_{\{s\}} g_{\{q\}} = 0, \;\;\;\;\; {\{s\}}\in {\{q\}}. 
\end{equation}
Imposing \ref{cluster_dec} first three eqautions in the system of chain equations take the next form:
\begin{equation}
i\hbar \frac{d}{dt} f_{a} - \left[  \hat{H}_{f},f_a  \right] = \sum_{f} \left[\langle \hat{V}_{af} \rangle_f, f_{a} \right]+\sum_{f} Tr_{f} \left[ \hat{V}_{af}, g_{af} \right],
\end{equation}
\begin{equation}
i\hbar \frac{d}{dt} f_{f} - \left[  \hat{H}_{f},f_f  \right] = \sum_{a} \left[\langle \hat{V}_{af} \rangle_a, f_{f} \right]+\sum_{a} Tr_{a} \left[ \hat{V}_{af}, g_{af} \right],
\end{equation}
\begin{eqnarray}
\nonumber i\hbar \frac{d}{dt} g_{af} + i\hbar f_f \frac{d}{dt} f_{a} + i\hbar f_a \frac{d}{dt} f_{f} - \left[ \hat{H}_{a} + \hat{H}_{f} +  \hat{V}_{af}, f_a f_f \right] = \\ \nonumber = \left[ \hat{H}_{a} + \hat{H}_{f} +  \hat{V}_{af}, g_{af} \right] +\sum_{a'\neq a} Tr_{a'} \left[ \hat{V}_{a'f}, f_a f_{a'} f_f \right] +\\ \nonumber + \sum_{a'\neq a} Tr_{a'} \left[ \hat{V}_{a'f}, f_{a}g_{a'f} \right] + \sum_{a'\neq a} Tr_{a'} \left[ \hat{V}_{a'f}, f_{a'}g_{af} \right] \\ + \sum_{a'\neq a} Tr_{a'} \left[ \hat{V}_{a'f}, f_{f}g_{aa'} \right] + \sum_{a'\neq a} Tr_{a'} \left[ \hat{V}_{a'f}, g_{aa'f} \right] +\\ \nonumber + \sum_{f' \neq f} Tr_{f'} \left[ \hat{V}_{af'}, f_{a} f_{f} f_{f'} \right] + \sum_{f' \neq f} Tr_{f'} \left[ \hat{V}_{af'}, f_{a} g_{ff'} \right] \\ \nonumber  + \sum_{f' \neq f} Tr_{f'} \left[ \hat{V}_{af'}, f_{f} g_{af'} \right] + \sum_{f' \neq f} Tr_{f'} \left[ \hat{V}_{af'}, f_{f'} g_{af} \right] + \\ \nonumber  + \sum_{f' \neq f} Tr_{f'} \left[ \hat{V}_{af'}, g_{aff'} \right] .
\end{eqnarray}
Here the following notation for mean value of operator was introduced: $\langle \hat{O} \rangle_f = Tr_f {\hat{O} \rho_f} $. In order to derive this equation in its final form one need to subtract equation for $f_a$  multiplied by $f_f$ and subtract equation for $f_f$ multiplied by $f_a$. In the result the dynamical equation for $g_{aa'f}$ is the next:
\begin{eqnarray} \label{g_af}
\nonumber i\hbar \frac{d}{dt} g_{af} - \left[ \hat{V}_{af}, f_a f_f \right] - \left[ \hat{H}_{a} + \hat{H}_{f} +  \hat{V}_{af}, g_{af} \right] = \\ \nonumber = - f_a  Tr_{a} \left[ \hat{V}_{af}, f_a f_f \right] - f_f  Tr_{f} \left[ \hat{V}_{af}, f_{a} f_{f}  \right] - \\ \nonumber - f_a  Tr_{a} \left[ \hat{V}_{af}, g_{af} \right] - f_f  Tr_{f} \left[ \hat{V}_{af}, g_{af}  \right] + \\ \nonumber +  \sum_{a'\neq a} Tr_{a'} \left[ \hat{V}_{a'f}, f_{a'}g_{af} \right] + \sum_{a'\neq a} Tr_{a'} \left[ \hat{V}_{a'f}, f_{f}g_{aa'} \right] + \\ \nonumber + \sum_{a'\neq a} Tr_{a'} \left[ \hat{V}_{a'f}, g_{aa'f} \right] + \sum_{f' \neq f} Tr_{f'} \left[ \hat{V}_{af'}, f_{a} g_{ff'} \right] \\ \nonumber  + \sum_{f' \neq f} Tr_{f'} \left[ \hat{V}_{af'}, f_{f'} g_{af} \right] + \sum_{f' \neq f} Tr_{f'} \left[ \hat{V}_{af'}, g_{aff'} \right] .
\end{eqnarray}
In this way one can obtain equations of motion for correlation matrices of $n$-particles. This procedure becomes more labor-intensive as the number of particles described by the correlation matrix increases. This procedure can be completely algorithmized, which will be shown further in the work. Further use of Bogolyubov chains is the following. To describe the properties of quantum emitters, it is necessary to find the master equation, which is represented by the equation for $f_a$. Formal solutions for density matrices and correlation operators are substituted into equation for $f_a$ and then using the necessary approximation one can obtain an equation with the desired degree of accuracy. The scheme of this system is presented in Fig. \ref{fig_1}.

\section{Structure of Classes and Their Methods}
\begin{figure}[t]
\centering
\includegraphics[scale=0.7]{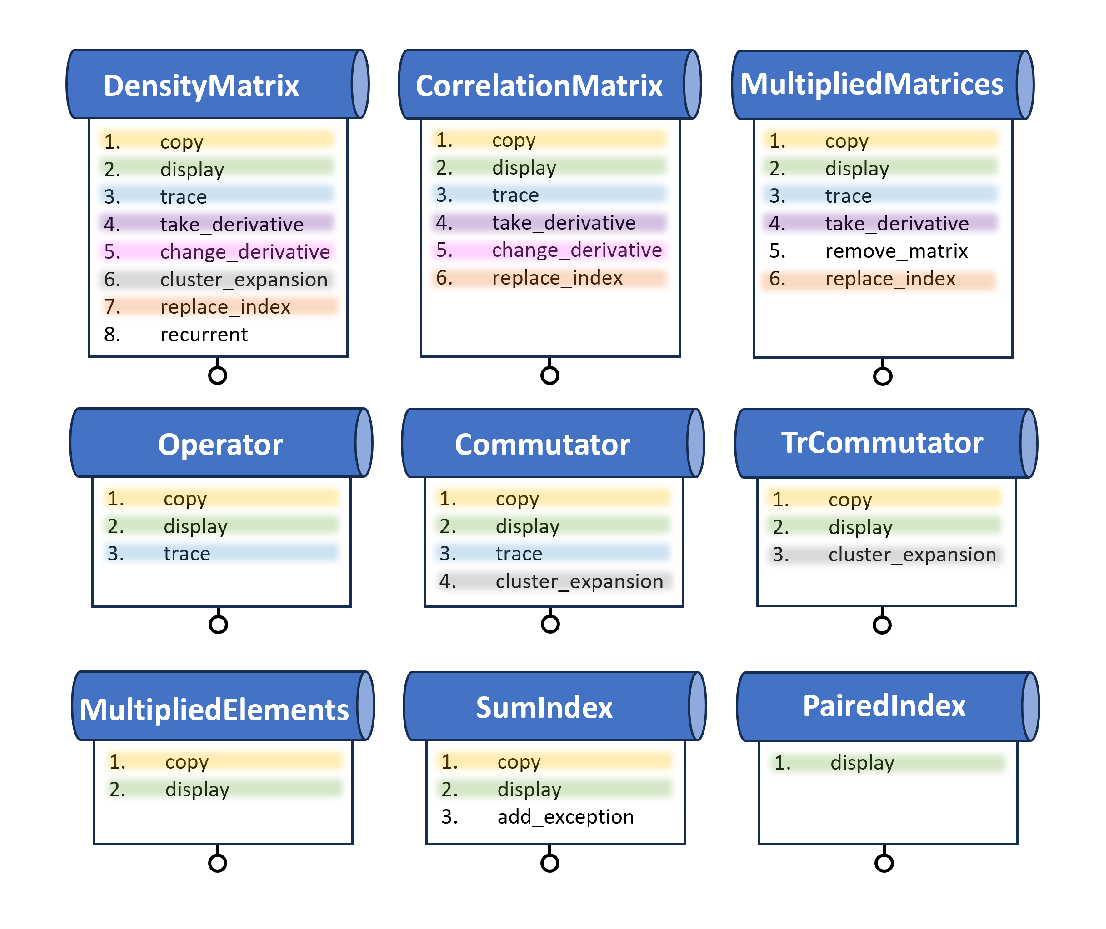}
\caption{The main classes used in the program and their methods. Common methods of classes in one color.}\label{fig_2}
\end{figure}

\begin{figure}[t]
\centering
\includegraphics[scale=0.5]{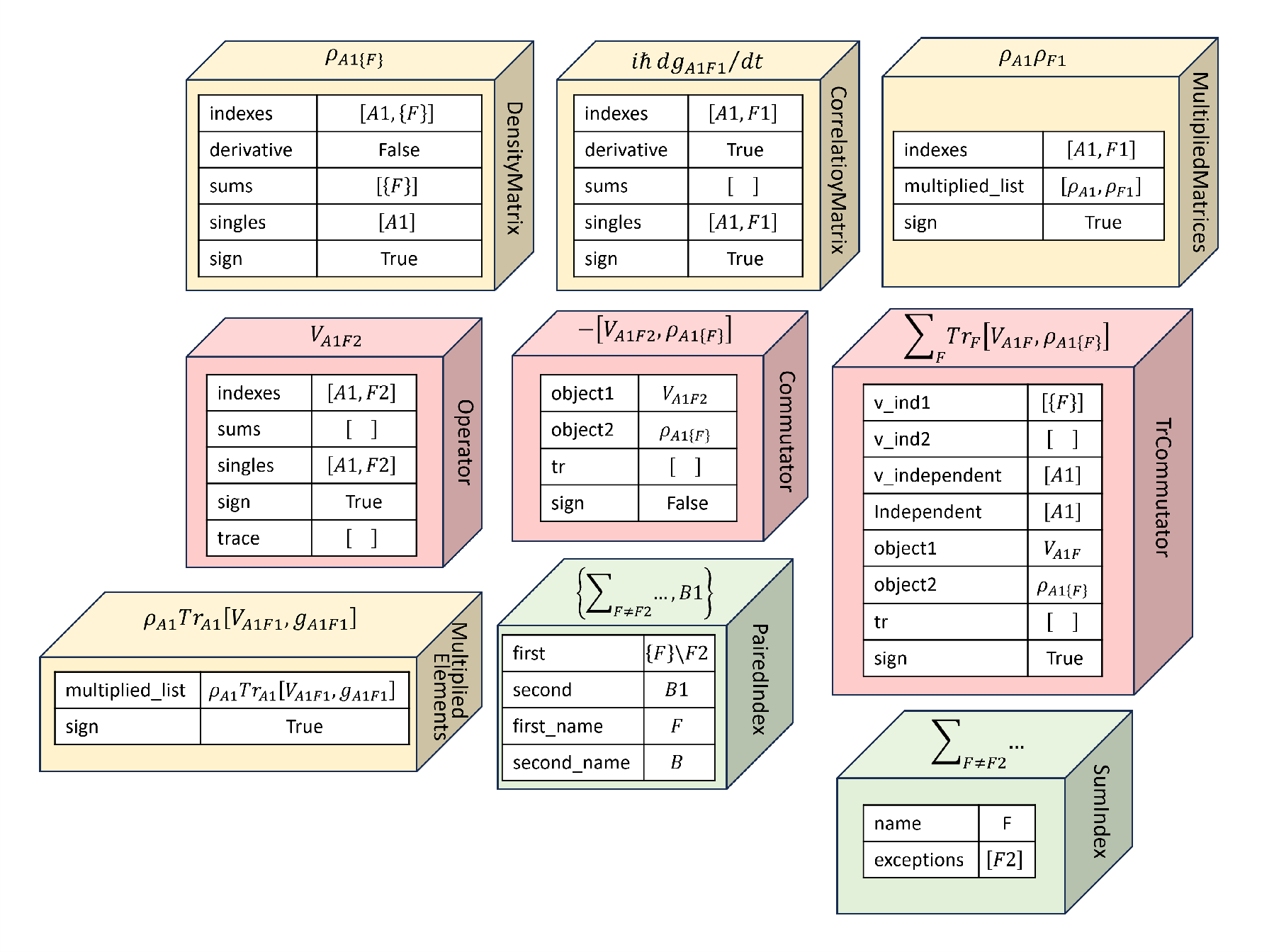}
\caption{The main classes used in the program and their attributes.}\label{fig_3}
\end{figure}

For programmatic generation of equations of BBGKY chains on the Python platform, it is necessary to create several special classes that describe all formal elements in the equations. The following classes were created in the implemented program code: DensityMatrix, CorrelationMatrix, MultipliedMatrices, Operator, Commutator, TrCommutator, MultipliedElements, SumIndex, PairedIndex, Zero, Identity. Figures \ref{fig_2} and \ref{fig_3} show the main classes with their methods and attributes, except for the Zero and Identity classes, which are auxiliary. The use of auxiliary classes is required for outputs in the case where the returned result corresponds to zero or one. 

Let us provide the necessity of creating each class and the sufficiency of the proposed set of classes. The system of particles can be described using the full density matrix $\rho$ of all particles, the evolution of which is subject to the von Neumann equation in the interaction picture \citep{scully_QO}, \citep{gerry2023introductory}:
\begin{equation} \label{vN_int}
i\hbar \frac{d}{dt} \rho = \left[ \hat{V}, \rho \right],
\end{equation}
where $\hat{V}$ is the interaction operator, where one can define interactions between particles or subsystems according to the model. Equation (\ref{vN_int}) is the von Neumann equation (\ref{vN_sch}) in the interaction picture. The description of systems with an infinite number of degrees of freedom (the number of particles or subsystems with their own degrees of freedom) is essential for light-matter interaction problems. The implementation of such a possibility is critically necessary for the correct description of the quantized electromagnetic field, since the filed modes represent an infinite set of one-type "particles". This also concerns phonon modes (temperature effects) or a continuous medium, which may fix the spatial location of the researchers. The introduction of such a continuous host medium can be formally implemented through a finite or infinite number of particles of the medium with certain model properties. To work with an infinite number of particles, the SumIndex class was introduced, the objects of which have two attributes: a name and exceptions from the sum. It is necessary to work with exceptions from the sum when for example one or more particles from an ensemble of similar particles are considered. Another example is derivation of an equation for one specific mode of the electromagnetic field. Equation (\ref{g_af}) from the previous section contains incomplete sums over the modes of quantized electromagnetic field, as well as over the emitters of the medium. If the "particle" is presented in the complete system in a single form, then it can be specified through a python build-in string class in the format "A1".
Objects of the DensityMatrix class represent density matrices of a finite or infinite number of "particles" (subsystems of the total ensemble). To implement subsystems of an infinite number of "particles", an object of the SumIndex class is used. This is an important property of the DensityMatrix class, since the density matrix can describe infinite ensembles of "particles" - a continuous medium, an electromagnetic field, phonons. Thus, it seems reasonable to write the short notation of the density matrix $\rho$ in a more complete form, explicitly indicating the subsystems that make up the complete system. For example, if the system consists of three emitters $A1$, $A2$, $A3$, embedded in a medium $\{B\}$ and interacting with the modes of the electromagnetic field $\{F\}$, then the density matrix in expanded form looks like this:
\begin{equation} \label{example_DM}
\rho = \rho_{A1A2A3\{B\}\{F\}}.
\end{equation}
The interaction between particles is specified through binary interaction potentials $\hat{V}_{uv}$, where $u$ and $v$ indicate specific "particles" between which the interaction occurs. In the program, the operators describing the interaction are represented through objects of the Operator class, whose attributes contain two indices ($u$ and $v$). The indices of the operators can be specified either through the SumIndex class or through a string index. The case where two or one indices are sum indices corresponds to the description of the interaction of two infinite ensembles of "particles" (for example, electromagnetic field modes and a material medium) or one infinite ensemble of similar "particles" and a single particle (for example, electromagnetic field modes and an emitter):
\begin{equation} \label{example_V}
\sum_{F} \sum_{B} \hat{V}_{FB} \;\;\;\;\; or \;\;\;\;\; \sum_{F} \hat{V}_{A1F}. 
\end{equation}
The general operator $\hat{V}$ consists of all specified interactions in the system. In the further derivation of the hierarchy of equations, operators with different indices are subtracted from each other. For this procedure, a special class PairedIndex was created, the object of which is a pair of indices (string, sum or mixed).

For the already given example (\ref{example_DM}) of interaction let each of the three emitters $A1$, $A2$, $A3$ interact with the modes of the electromagnetic field $\{F\}$, and let each of the emitters of the medium $\{B\}$ also interact with the modes of the electromagnetic field $\{F\}$. For such system equation (\ref{vN_int}) with explicit interaction  (\ref{example_V}) takes the following form:
\begin{eqnarray}
i\hbar \frac{d}{dt} \rho_{A1A2A3\{B\}\{F\}} =  \\ \nonumber \left[ \sum_{F} \hat{V}_{A1F} + \sum_{F} \hat{V}_{A2F} + \sum_{F} \hat{V}_{A3F} + \sum_{F} \sum_{B} \hat{V}_{FB},  \rho_{A1A2A3\{B\}\{F\}} \right].
\end{eqnarray}
Since the commutator is linear, one can rewrite the last equation in sums of commutators:
\begin{eqnarray} \label{example_vN}
\nonumber i\hbar \frac{d}{dt} \rho_{A1A2A3\{B\}\{F\}} =   \left[ \sum_{F} \hat{V}_{A1F},  \rho_{A1A2A3\{B\}\{F\}} \right]+ \\ + \left[ \sum_{F} \hat{V}_{A2F},  \rho_{A1A2A3\{B\}\{F\}} \right] + \left[\sum_{F} \hat{V}_{A3F},  \rho_{A1A2A3\{B\}\{F\}} \right]+ \\ \nonumber + \left[\sum_{F} \sum_{B} \hat{V}_{FB},  \rho_{A1A2A3\{B\}\{F\}} \right].
\end{eqnarray}
To describe the commutators the Commutator and TrCommutator classes were introduced, the attributes of which consist of an object in the first place (the interaction operator) and an object in the second place (the density matrix, the correlation matrix, the product of the density matrices and/or the correlation matrices). The TrCommutator class differs from the Commutator class in that it can have an index that simultaneously refers to both the trace and the operator. An example of such a commutator is shown in Figure \ref{fig_3} in the corresponding cell. The use of this class will be required at the following stages of deriving the BBGKY hierarchy of equations.
As it was previously described, after the system (the set of particles and the interactions between them) is specified, the operation of taking a trace is applied to equation (\ref{example_vN}). As a result, one obtains equations for the reduced density matrices. In the example under consideration, in order to obtain the density matrix of three particles $\rho_{A1F1B1}$ one needs to take a trace over all particles that are not present in the desired density matrix: $A2$, $A3$, $\{F\}\backslash F1$, $\{B\}\backslash B1$.
The next step in the derivation is to use the cluster decomposition of the density matrix with finite indices. For example, for $\rho_{A1F1B1}$ the cluster decomposition is:
\begin{equation}
\rho_{A1F1B1} = \rho_{A1}\rho_{F1} \rho_{B1} + \rho_{A1}g_{F1B1}+ \rho_{F1}g_{A1B1}+\rho_{B1}g_{A1F1}+ g_{A1F1B1},
\end{equation}
where $g_{...}$ are correlation matrices. In this equation two-particle and three-particle correlation matrices are presented. In the program, objects of the CorrelationMatrix class are various correlation matrices of a finite number $n$ of "particles". It is worth noting that, unlike density matrices, correlation matrices can have only finite indices. The terms in the expansion are a product of density matrices, products of density matrices and correlation matrices, and a single correlation matrix. There first 4 terms are represented in the program as objects of MultipliedMatrices class. The objects of MultipliedMatrices class contain a list of references to objects of the DensityMatrix or CorrelationMatrix classes for which the product takes place. A similar class MultipliedElements is a product of various objects of other classes (an example is shown in Figure \ref{fig_3}).

To work with class objects that represent elements of the BBGKY chains it is necessary to introduce methods. Methods are shown in Figure \ref{fig_2}. Methods that can be similarly applied to objects of different classes have the same names and are highlighted in the same color. The copy(self) method is designed to create an identical copy of a class object and it is presented in each class, except for the PairedIndex, Identity, and Zero classes. The display(self) method is used to output (as a string) the class object and if is presented in the methods of each class. Examples of the method display(self) are shown in Table \ref{tab1}. The trace(self, indexes) method is presented in the DensityMatrix, CorrelationMatrix, MultipliedMatrices, Operator, and Commutator classes. Here "indexes" is a list of indices by which the trace of operators (matrices) is taken. This list can contain indices of any type. The trace(self, indexes) method operates on objects of the DensityMatrix class as follows: from the class object's own indexes all the method trace "indexes" are removed. If the object's own indexes still remain, the method returns the class object with the remaining indexes. Otherwise, the method returns one. The general procedure of taking trace is presented in Table \ref{tab2}. For the CorrelationMatrix class, the situation is the opposite: if at least one element from trace method "indexes" is in the indexes of the CorrelationMatrix object, the output is Zero class object. This follows from constrain (\ref{corr_mat_constrain}) imposed by cluster decomposition on the correlation matrix. In the case of the Operator class object, if the intersection of the object's indexes and indexes entered in the trace(self, indexes) method is nonempty, the intersecting indexes are added to the object's attribute self.tr.
\begin{table}[t]
\setlength{\extrarowheight}{15pt}
\centering
\begin{tabular}{| m{2cm} | m{6cm} | m{4cm} |}
\hline
\multicolumn{1}{|c|}{\textbf {Mathematical expression}} & \multicolumn{1}{|c|}{\textbf {Code}} & \multicolumn{1}{|c|}{\textbf {Result of self.display }}\\ \hline  
\multicolumn{1}{|c|}{$0$}  &  \multicolumn{1}{|c|}{Zero()} & \multicolumn{1}{|c|}{0} \\  \hline
\multicolumn{1}{|c|}{$1$} & \multicolumn{1}{|c|}{Identity()} & \multicolumn{1}{|c|}{1} \\ \hline  
\multicolumn{1}{|c|}{$\sum_{F / F_1}$}  & \multicolumn{1}{|c|}{SumIndex('F',['F1'])} & \multicolumn{1}{|c|}{sum\_\{F\}/F1} \\  \hline
\multicolumn{1}{|c|}{$\left( A1, \lbrace F \rbrace \right)$} & \multicolumn{1}{|c|}{PairedIndex('A1', SumIndex('F'))} & \multicolumn{1}{|c|}{\{A1, sum\_\{F\}\} } \\ \hline  
\multicolumn{1}{|c|}{$\rho_{A_1 \lbrace F \rbrace}$} &  DensityMatrix (False, ['A1', SumIndex('F')]) & \multicolumn{1}{|c|}{rho\_A1\{F\}} \\ \hline
\multicolumn{1}{|c|}{$i\hbar \frac{\partial}{\partial t} g_{A_1F_1}$ }& CorrelationMatrix (True, ['A1', 'F1']) & \multicolumn{1}{|c|}{i hbar d/dt g\_A1F1} \\   \hline
\multicolumn{1}{|c|}{$\rho_{A_1}g_{A_2F_1}$ }& MultipliedMatrices (DensityMatrix (False, ['A1']), CorrelationMatrix (False, ['A2', 'F1']) ) & \multicolumn{1}{|c|}{rho\_A1 * g\_A2F1} \\  \hline
\multicolumn{1}{|c|}{$\sum_F \hat{V}_{A_1 F}$ }& Operator(['A1', SumIndex('F')]) & \multicolumn{1}{|c|}{sum\_\{F\} V\_A1F} \\ \hline
\multicolumn{1}{|c|}{$\left[ \hat{V}_{A_1B_1}, g_{A_1B_1} \right]$} & Commutator(Operator(['A1', 'B1']), CorrelationMatrix(False, ['A1', 'B1'])) & \multicolumn{1}{|c|}{ [ V\_A1B1, g\_A1B1 ]} \\ \hline
\multicolumn{1}{|c|}{$\sum_F Tr_F \left[ \hat{V}_{A1F}, g_{A1F} \right] $} & TrCommutator([SumIndex('F')], [], 'A1', CorrelationMatrix(False, ['F', 'A1'])) & sum\_\{F\} Tr\_F [V\_A1F, g\_A1F] \\ \hline 
\multicolumn{1}{|c|}{$\rho_{F_1}Tr_{F_1} \left[ \hat{V}_{A_1F_1}, \rho_{A_1F_1} \right]$} &  MultipliedElements( [DensityMatrix(False, ['F1']), TrCommutator(['F1'], [], ['A1'], DensityMatrix(False, ['A1', 'F1']))]) &  rho\_F1 * Tr\_F1 [V\_['A1']F1, rho\_A1F1] \\  
  \hline
\end{tabular}
\caption{Examples of self.display for all types of classes.}\label{tab1}
\end{table}
\begin{table}[t]
\centering
\begin{tabular}{| c | c |}
\hline
\multicolumn{2}{| c|}{$Tr_{SQ} \rho_{QP}$} \\ \hline
  $P = \emptyset$ & $Tr_{SQ} \rho_{QP}=1$ \\ \hline
  $P \neq \emptyset $ & $Tr_{SQ} \rho_{QP}=\rho_{p}$ \\ \hline
\end{tabular}
\caption{The procedure of taking trace in general.}\label{tab2}
\end{table}
For objects of the MultipliedMatrices class, the trace(self, indexes) method works as follows. The self.multiplied\_list class attribute contains references to objects of the DensityMatrix and CorrelationMatrix classes. The trace(self, indexes) method is applied to each element of self.multiplied\_list and returns one of the following results:
\begin{enumerate}
        \item A new object of the MultipliedMatrices class, if among the returned objects there is no Zero class object, while elements of which the trace operation returns Identity are not added to self.multiplied\_list.
        \item An object of the Zero class, if at least one of the returned elements is an object of the Zero class.
        \item An object of the Identity class, if the trace(self, indexes) function returned an object of the Identity class for each of the elements of self.multiplied\_list.
        \item An object of the CorrelationMatrix or DensityMatrix class, if the length of self.multiplied\_list excluding objects of the Identity class is equal to one.
\end{enumerate}
Examples of applying the trace(self, indexes) method to objects of the MultipliedMatrices class are given in Table \ref{tab3}.
\begin{table}[t]
\centering
\begin{tabular}{| c | c |}
\hline
  \textbf{Example} & \textbf{Class of the output object} \\ \hline
  $Tr_{A_1}(\rho_{A_1}\rho_{A_2}\rho_{A_3}) = \rho_{A_2}\rho_{A_3} $ & MultipliedMatrices  \\ \hline
  $Tr_{A_1 A_2}(\rho_{A_1}\rho_{A_2}\rho_{A_3}) = \rho_{A_3} $&  DensityMatrix \\ \hline
  $Tr_{A_1 A_2 A_3}(\rho_{A_1}\rho_{A_2}\rho_{A_3}) = 1$ & Identity \\ \hline
  $Tr_{A_2}(\rho_{A_1}g_{A_2 A_3} = 0$ & Zero \\ \hline
\end{tabular}
\caption{Examples of trace taking procedure of objects of  MultipliedMatrices class.}\label{tab3}
\end{table}

\begin{figure}[t]
\centering
\includegraphics[scale=0.5]{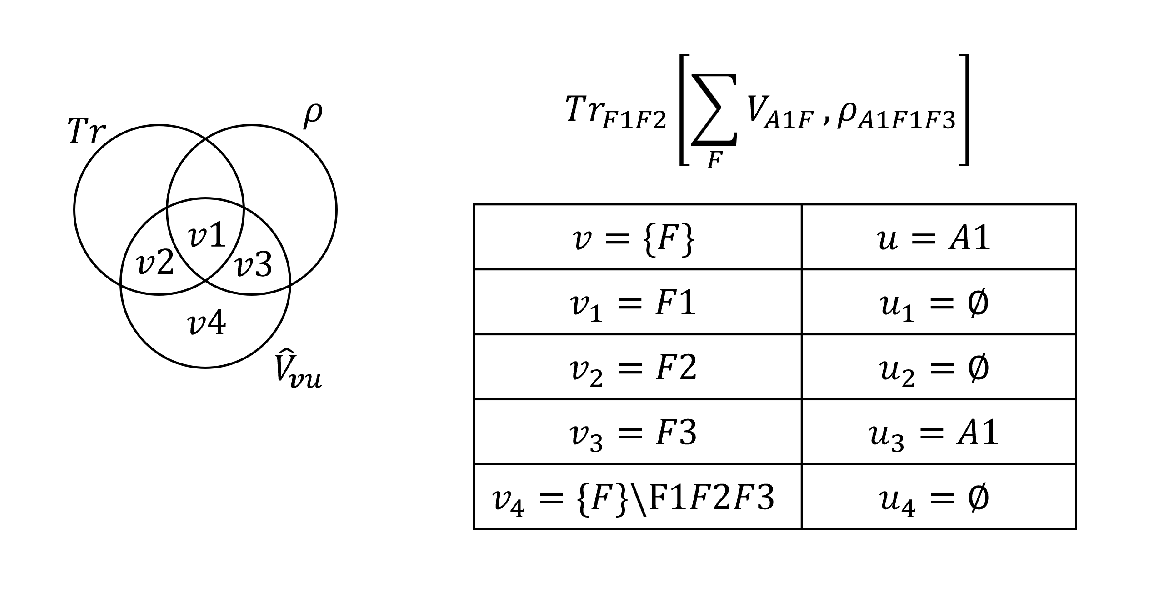}
\caption{Scheme (left) and example (right) of segregation operator indices into a group.}\label{fig_4}
\end{figure}

\begin{figure}[t]
\centering
\includegraphics[scale=0.5]{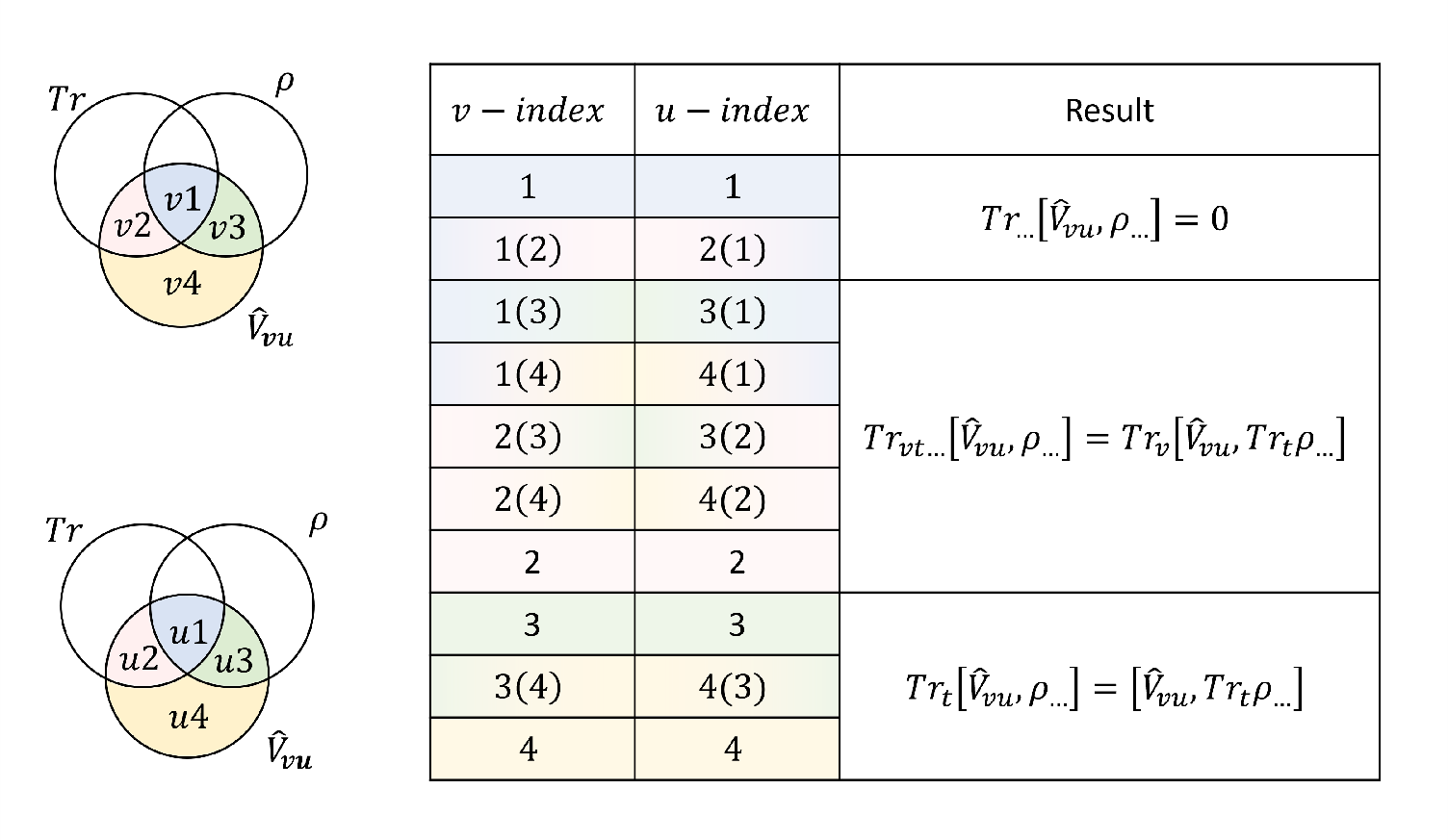}
\caption{Results of applying the trace method to a Commutator class object based on indices segregation into groups.}\label{fig_5}
\end{figure}

For the Commutator class, the trace(self, indexes) method works as follows. Let $v$-index denote the first index of the Operator class object (self.object1), and $u$-index the second. If the index is single, it belongs to one of the four groups ($v_1$, $v_2$, $v_3$, and $v_4$). If the index is SumIndex class object, it can be divided into these groups. The sequence of segregation indices into groups and an example are shown in Figure \ref{fig_4}. After dividing indices of Commutator class object into four groups, for each pair of groups (if they are not empty), a Zero, Commutator or TrCommutator class object is returned according to Figure \ref{fig_5}.

\begin{figure}[t]
\centering
\includegraphics[scale=0.55]{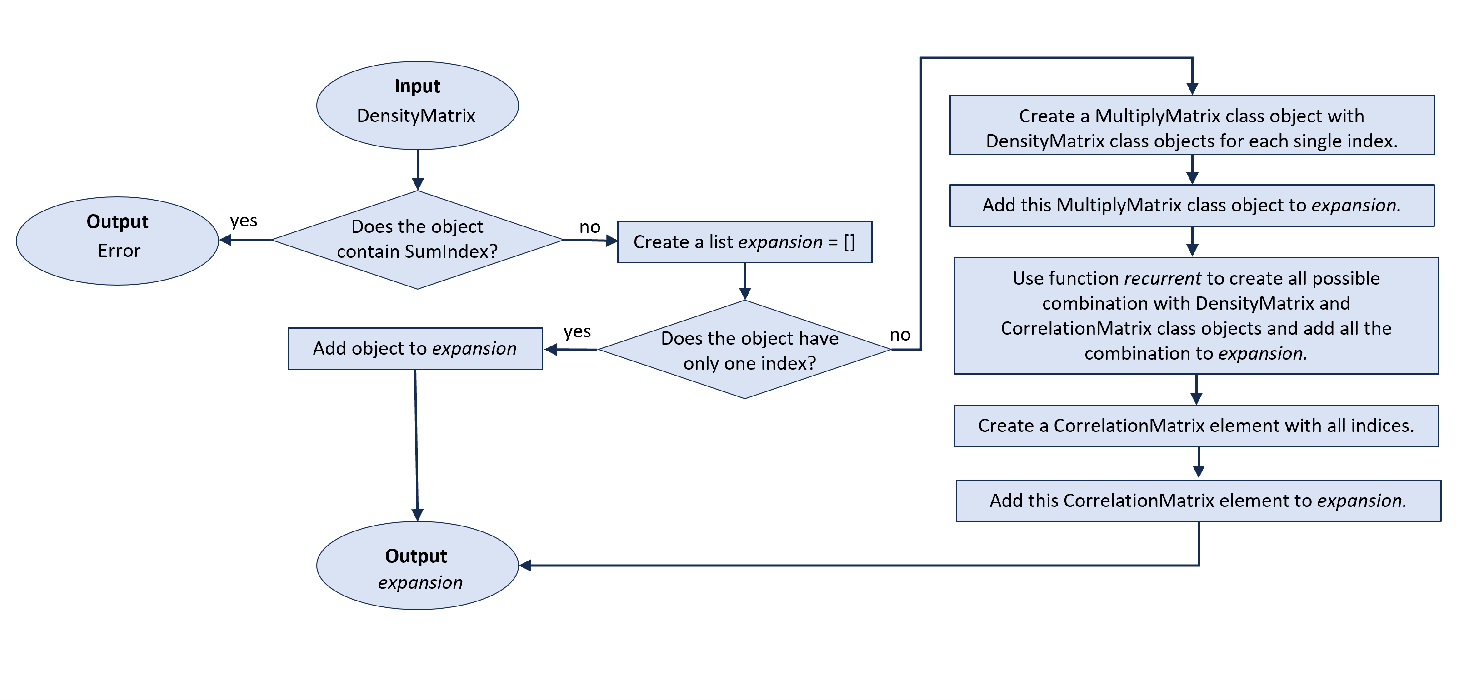}
\caption{Illustration of the main part of the algorithm for cluster decomposition (cluster\_expansion(self) method) of a multi-component density matrix (DensityMatrix class object).}\label{fig_6}
\end{figure}

It is worth noting that there is no need to introduce the trace(self, indexes) method for the TrCommutator class, since the current algorithm for deriving BBGKY chains allows one to avoid taking trace from a commutator, from which trace has already been taken. In addition, if such a need arises, one can take a "step back" and by re-setting the initial commutator, take trace by the set of indices required from the trace(self, indexes) method and the indices that were in attribute of the TrCommutator class object under consideration. 
The main part of the algorithm of the cluster\_expansion(self) method for the DensityMatrix class object is schematically shown in Figure \ref{fig_6}. The method returns a list of elements (objects of the MultipliedMatrixes or CorreleationMatrix classes) formed as follows. First an MultipliedMartix class object composed of DensityMatrix class objects with single index for each index of the original density matrix is added to the beginning of the list. Then, according to the number of iterations, elements are formed, containing several DensityMatrix class objects  and one CorrelationMatrix class object with the remaining indices. At the end of the list an object of the CorrelationMatrix class is added, which is the final correlation matrix in the cluster decomposition with all the original indices.

The cluster\_expansion(self) method for the Commutator and TrCommutator classes works in such a way that the argument self.object2 in these classes is subject to the action of cluster\_expansion(self). The output is a list of elements of the Commutator or TrCommutator classes, where self.object2 is an element from the list obtained after applying the cluster\_expansion(self) method to self.object2. The remaining methods of the classes are auxiliary methods to the above. They do not need a detailed description.

In order to obtain the equation for the correlation matrix with the given indices (required\_indexes), the following scheme is used:
\begin{enumerate}
        \item According to the input sets of subsystems and the given interactions between them an equation is constructed in the form of two lists (left hand side and right hand side). Each side contains class objects as equation elements. For example, for the equation $i\hbar \frac{d}{dt} \rho= \left[ \sum_F \hat{V}_{A1F}, \rho \right]+\left[ \sum_F \sum_B \hat{V}_{BF}, \rho \right]$ the left hand side (lhs\_main) will contain one element representing the density matrix of all "particles" in the system, and the right hand side (rhs\_main) will contain two elements represented by objects of the Commutator class.
        \item The self.trace and self.cluster\_expansion methods are applied to the elements of the lists. New lists lhs\_main rhs\_main are formed from the output of methods operations.
        \item 3. An object of the DensityMatrix class is created with the specified indices (required\_indexes). The self.cluster\_expansion method is applied to this object, the output is written to the list. Then, the self.take\_derivative method is applied to each element of the list, the result is written to the derivative\_list.
        \item The loop through all elements of the derivative\_list except for the last one is applied. Such elements are objects of the MultipliedMatrixes class. For the element with the self.derivative = True in self.multiplied\_list a kinetic equation is generated. If this element is an object of the DensityMatrix class, then the generation occurs by taking the required trace from the original equation. Another possible case is that the element is an object of the CorrelationMatrix class, then this algorithm is applied recurrently.
        \item The resulting kinetic equation, which elements are also written in left hand side and right hand side list, is then multiplied by the remaining elements in self.multiplied\_list element. The result is subtracted from lhs\_main and rhs\_main, respectively. To subtract elements functions for comparing and deleting elements from lists were written specially.
\end{enumerate}

\section{Comparison with analytical derivation}

\begin{figure}[t]
\centering
\includegraphics[scale=0.75]{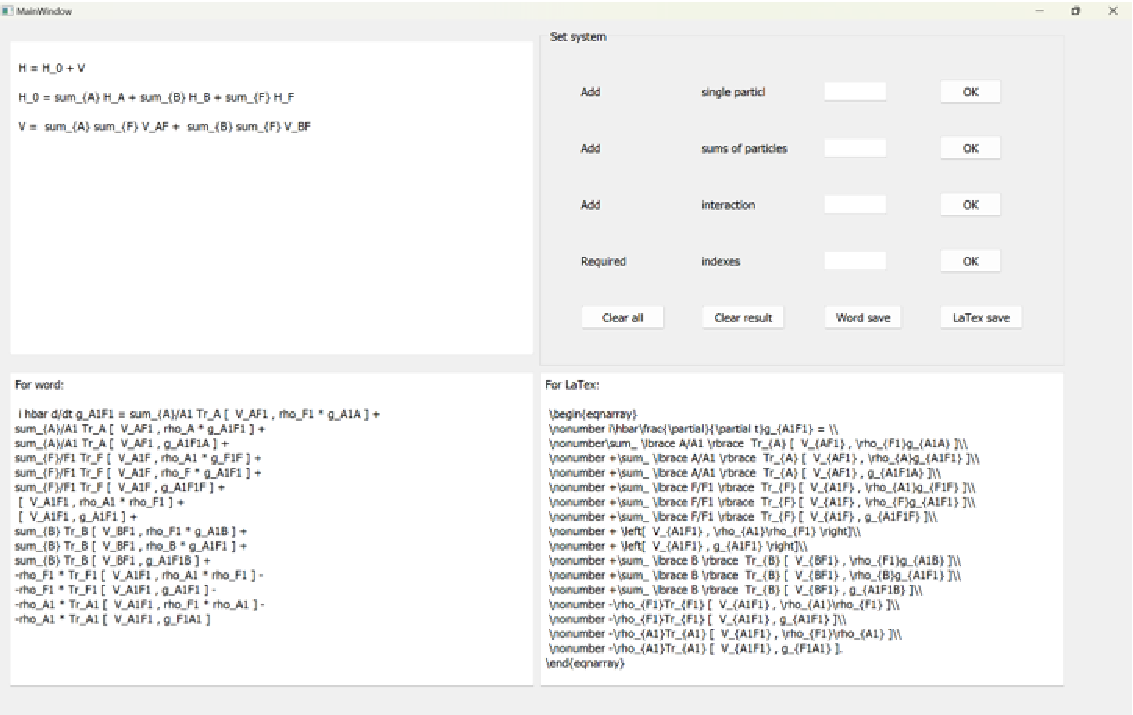}
\caption{Program interface and an example of its operation.}\label{fig_7}
\end{figure}

In this section, we compare the results of the software derivation of the BBGKY chain of equations with its analytical derivations given in works \citep{kuznetsov2011_TMP}, \citep{kuznetsov2011_JETP}, \citep{lozing2020} (supplementary material). In these works, the photoluminescence of quantum emitters embedded in a transparent medium was considered. The authors derived BBGKY chains of equation for a statistical ensemble of three types of particles. The ensemble included a set of quantum emitters with two states, a set of particles that form a transparent host medium, and a continuum of photon modes. The number of equation of the BBGKY quantum chain necessary for the correct consideration of the problem was limited by equations for second-order correlation matrices. The created program was tested for constructing equations for all correlation matrices that represent correlations between two different subsystems of the same type and correlations between subsystems of different types. Figure \ref{fig_7} shows the program interface where the system under consideration is specified by entering the energy operators. The energy operator is specified by reading the names of single subsystems (e.g., 'A1', 'B3', etc.), and/or the names of subsystems, the number of which is actually or formally infinite (e.g., 'A', 'B', or 'F'). Then, the interaction between the named subsystems is indicated (e.g., 'A1F' or 'AF'). To compare with analytical derivation from the article two types of subsystems without interaction between them were entered. These sets of subsystems corresponded to ensembles of quantum emitters ('A') and the set of medium particles ('B'). Then, the presence of subsystems of the third type was indicated, which corresponded to the continuum of quantized photon modes ('F'). Between this type of subsystems paired interactions with subsystems of the first and second types ('AF' and 'BF') were introduced, which corresponded to the physical picture of the interaction of material particles through the radiation field. The program code generated the correct energy operator, which is shown in the upper left window in Figure \ref{fig_7}. For generation the kinetic equation of the correlation matrix of the BBGKY quantum chain the required indices of the correlation matrix need to be specified. The specified indices are transferred to the program block, which operates according to the algorithm for constructing a correlation matrix described in the previous section. As a result, the program outputs an equation for the correlation matrix or density matrix, in the case of one specified index.
According to the described above system the program generated equations for single-particle density matrices that correspond to the emitter, the particle of the medium, and the mode of the quantized electromagnetic field:
 \begin{equation} \label{rho_A1}
i\hbar\frac{\partial}{\partial t}\rho_{A1} = 
 \sum_{F }  Tr_{F} [  V_{A1F} , \rho_{A1}\rho_{F} ] +\sum_{F}  Tr_{F} [  V_{A1F} , g_{A1F} ],
\end{equation} 
\begin{eqnarray} \label{rho_F1} 
 \nonumber i\hbar\frac{\partial}{\partial t}\rho_{F1} = \sum_A  Tr_{A} [  V_{AF1} , \rho_{F1}\rho_{A} ] +\sum_A  Tr_{A} [  V_{AF1} , g_{F1A} ]\\ 
  +\sum_B  Tr_{B} [  V_{BF1} , \rho_{F1}\rho_{B} ]+\sum_B Tr_{B} [  V_{BF1} , g_{F1B} ].
\end{eqnarray}

The equations are presented in formula form, corresponding to the LaTex syntax compiled for output. For the case, when there is a single specified index, the program produces equations for the density matrix, since for one particle subsystem the concept of correlation is lost. Kinetic equation for density matrix $\rho_{B1}$ is identical to kinetic equation for $\rho_{A1}$ with replaced index $A1$ to $B1$. Equations (\ref{rho_A1}) and (\ref{rho_F1}) are in fully agreement with equations (14) and (15) from \citep{kuznetsov2011_TMP}, equations (10) and (11) from \citep{kuznetsov2011_JETP}, equations for $\rho_e$ and $\rho_p$ form supplementary material of work \citep{lozing2020}. Below correlation matrices constructed by the program are presented, which correspond to different variants of correlation between particles considered in works \citep{kuznetsov2011_JETP}, \citep{kuznetsov2011_TMP}, \citep{lozing2020}.
\begin{eqnarray}  \label{g_A1A2}
 \nonumber i\hbar\frac{d}{dt}g_{A1A2} = \sum_F Tr_{F} [  V_{A1F} , \rho_{A1}g_{A2F} ] +\sum_F Tr_{F} [  V_{A1F} , \rho_{F}g_{A1A2} ]\\  +\sum_F Tr_{F} [  V_{A1F} , g_{A1A2F} ] +\sum_F Tr_{F} [  V_{A2F} , \rho_{A2}g_{A1F} ]\\ 
 \nonumber +\sum_F Tr_{F} [  V_{A2F} , \rho_{F}g_{A1A2} ]+\sum_F Tr_{F} [  V_{A2F} , g_{A1A2F} ],
\end{eqnarray}
\begin{eqnarray} \label{g_F1F2}
\nonumber i\hbar\frac{d}{dt}g_{F1F2} = \sum_A  Tr_{A} [  V_{AF1} , \rho_{F1}g_{F2A} ] +\sum_B Tr_{B} [  V_{BF1} , \rho_{F1}g_{F2B} ]\\ 
 \nonumber +\sum_A Tr_{A} [  V_{AF1} , \rho_{A}g_{F1F2} ]+\sum_A Tr_{A} [  V_{AF1} , g_{F1F2A} ]\\ 
 \nonumber +\sum_A Tr_{A} [  V_{AF2} , \rho_{F2}g_{F1A} ] +\sum_A Tr_{A} [  V_{AF2} , \rho_{A}g_{F1F2} ]\\ +\sum_A  Tr_{A} [  V_{AF2} , g_{F1F2A} ]+\sum_B Tr_{B} [  V_{BF1} , \rho_{B}g_{F1F2} ]\\ 
 \nonumber +\sum_B Tr_{B} [  V_{BF1} , g_{F1F2B} ] +\sum_B Tr_{B} [  V_{BF2} , \rho_{F2}g_{F1B} ]\\ 
 \nonumber +\sum_B Tr_{B} [  V_{BF2} , \rho_{B}g_{F1F2} ] +\sum_B Tr_{B} [  V_{BF2} , g_{F1F2B} ],
\end{eqnarray}
\begin{eqnarray}  \label{g_A1B1}
 \nonumber i\hbar\frac{d}{dt}g_{A1B1} = \sum_{F} Tr_{F} [  V_{A1F} , \rho_{A1}g_{B1F} ] +\sum_{F} Tr_{F} [  V_{A1F} , \rho_{F}g_{A1B1} ]\\ 
 +\sum_{F}  Tr_{F} [  V_{A1F} , g_{A1B1F} ] +\sum_{F}  Tr_{F} [  V_{B1F} , \rho_{B1}g_{A1F} ]\\ 
 \nonumber +\sum_{F}  Tr_{F} [  V_{B1F} , \rho_{F}g_{A1B1} ] +\sum_{F} Tr_{F} [  V_{B1F} , g_{A1B1F} ],
\end{eqnarray}
\begin{eqnarray} \label{g_A1F1}
\nonumber i\hbar\frac{d}{dt}g_{A1F1} = \sum_{A/A1}  Tr_{A} [  V_{AF1} , \rho_{F1}g_{A1A} ] + \sum_{A/A1}  Tr_{A} [  V_{AF1} , \rho_{A}g_{A1F1} ] \\ \nonumber + \sum_{A/A1}  Tr_{A} [  V_{AF1} , g_{A1F1A} ] +\sum_{ F/F1 }  Tr_{F} [  V_{A1F} , \rho_{A1}g_{F1F} ] \\ \nonumber + \sum_{F/F1}  Tr_{F} [  V_{A1F} , \rho_{F}g_{A1F1} ] +\sum_{F/F1}  Tr_{F} [  V_{A1F} , g_{A1F1F} ] \\ + \left[  V_{A1F1} , \rho_{A1}\rho_{F1} \right] + \left[  V_{A1F1} , g_{A1F1} \right] \\ 
 \nonumber + \sum_{B} Tr_{B} [  V_{BF1} , \rho_{F1}g_{A1B} ] +\sum_{B} Tr_{B} [  V_{BF1} , \rho_{B}g_{A1F1} ]  \\ 
 \nonumber +\sum_{B} Tr_{B} [  V_{BF1} , g_{A1F1B} ] -\rho_{F1}Tr_{F1} [  V_{A1F1} , \rho_{A1}\rho_{F1} ]\\ 
 \nonumber -\rho_{F1}Tr_{F1} [  V_{A1F1} , g_{A1F1} ]-\rho_{A1}Tr_{A1} [  V_{A1F1} , \rho_{F1}\rho_{A1} ]\\ 
 \nonumber -\rho_{A1}Tr_{A1} [  V_{A1F1} , g_{F1A1} ].
\end{eqnarray}
First pair of equations (\ref{g_A1A2}) and (\ref{g_F1F2}) describes correlation between "particles" of one type: emitter - emitter and phonon mode - phonon mode correlations. These equations were derived in work \citep{kuznetsov2011_TMP} (see equations (25) and (23)). Equations (\ref{g_A1A2}) and (\ref{g_F1F2} include all terns presented in work \citep{kuznetsov2011_TMP} as well as correlations of higher order (correlation matrices of three particles). Equations (\ref{g_A1B1}) and (\ref{g_A1F1}) were derived in all works under consideration: equations (24) and (22) in \citep{kuznetsov2011_TMP}, equations (12) and (13) in \citep{kuznetsov2011_JETP}, and equation for $g_{eh}$ and $g_{ep}$ in \citep{lozing2020}. In works under consideration the authors limited themselves to second-order correlation with the equation for $g_{A1B1}$. However, for $g_{A1F1}$ not only this, but other rather subtle approximations were applied for terms with second-order correlation matrices. The systems of equations presented above include both the terms presented in the articles and those that the authors rejected within the framework of the problem under consideration.

\section{Conclusion}

The program was tested on a PC to evaluate the time needed for constriction of the required correlation matrix using technical resources available on average. The program was tested on a PC with the following system parameters:
\begin{itemize}
        \item[] Processor: 11th Gen Intel(R) Core(TM) i5-1135G7 @ 2.40GHz 2.42 GHz,
        \item[] RAM: 16.0 GB (available: 15.8 GB),
        \item[] System type: 64-bit operating system, x64 processor.
\end{itemize}
The time to find a given correlation matrix depends both on the given system and the combination of indices that characterize the correlation matrix. To demonstrate this two systems were considered. System 1 was a single emitter ('A1') and a set of modes of the quantized electromagnetic field ('F'), between which there was an interaction ('A1F'). The Hamiltonian of system 1 is:
\begin{equation}
\hat{H}_1 = \hat{H}_{A1} + \sum_{F} \hat{H}_F + \sum_{F} \hat{V}_{A1F}.
\end{equation}
System 2 is system 1 supplemented with medium particles ('B'), which also interact with the modes of the quantized electromagnetic field. The Hamiltonian of system 2 is the following form:
\begin{equation}
\hat{H}_2 = \hat{H}_{A1} + \sum_{F} \hat{H}_F + \sum_{B} \hat{H}_B + \sum_{F} \hat{V}_{A1F} + \sum_{F} \sum_{B}\hat{V}_{BF}.
\end{equation}
Tables \ref{tab4} and \ref{tab5} present average times for finding the correlation matrices of the third and fourth orders for systems 1 and 2 respectively. 
It is shown that all equations up to the third order of the BBGKY chain completely coincide with the result of the analytical derivation. The equations for the fourth  order for these systems are constructed by the program for the first time and are unique.

\begin{table}[t]
\centering
\begin{tabular}{| c | c | c | c |}
\hline
\multicolumn{2}{| p{6cm}|}{\textbf {Average time to obtain three-particle correlation matrices}} & \multicolumn{2}{p{6cm}| }{\textbf {Average time to obtain four-particle correlation matrices}} \\  \hline
  $g_{A_1 F_1 F_2}$ & $0.160 \pm 0.003$, sec & $g_{A_1 F_1 F_2 F_3}$ & $7.816 \pm 1.242$, sec \\ \hline
  $g_{F_1 F_2 F_3}$ & $0.076 \pm 0.002$, sec & $g_{F_1 F_2 F_3 F_4}$ & $1.780 \pm 0.248$, sec  \\
  \hline
\end{tabular}
\caption{Program execution time for derivation of third-order and fourth-order correlation matrices of system 1.}\label{tab4}
\end{table}

\begin{table}[t]
\centering
\begin{tabular}{| c | c | c | c |}
\hline
\multicolumn{2}{| p{6cm}|}{\textbf {Average time to obtain three-particle correlation matrices}} & \multicolumn{2}{p{6cm}| }{\textbf{Average time to obtain four-particle correlation matrices}} \\  \hline
  $g_{F_1 F_2 F_3}$ & $0.470 \pm 0.006$, sec & $g_{F_1 F_2 F_3 F_4}$ & $8.378 \pm 1.860$, sec \\ \hline
  $g_{F_1 F_2 B_1}$ & $0.729 \pm 0.008$, sec & $g_{F_1 F_2 F_3 B_1}$ & $34.988 \pm 4.047$, sec  \\ \hline
  $g_{F_1 B_1 B_2}$ & $0.587 \pm 0.004$, sec & $g_{F_1 F_2 B_1 B_2}$ & $48.277 \pm 4.268$, sec \\ \hline
  $g_{B_1 B_2 B_3}$ & $0.134 \pm 0.002$, sec & $g_{F_1 B_1 B_2 B_3}$ & $28.277 \pm 3.264$, sec  \\ \hline
  $g_{A_1 F_1 F_2}$ & $0.470 \pm 0.004$, sec & $g_{B_1 B_2 B_3 B_4}$ & $2.886 \pm 0.121$, sec \\ \hline
  $g_{A_1 F_1 B_1}$ & $0.471 \pm 0.008$, sec & $g_{A_1 F_1 F_2 F_3}$ & $17.808 \pm 0.618$, sec  \\ \hline
  $g_{A_1 B_1 B_2}$ & $0.119 \pm 0.002$, sec & $g_{A_1 F_1 F_2 B_1}$ & $38.572 \pm 5.386$, sec \\ \hline
  - & - & $g_{A_1 F_1 B_1 B_2}$ & $26.817 \pm 0.784$, sec  \\ \hline
  - & - & $g_{A_1 B_1 B_2 B_3}$ & $3.071 \pm 0.217$, sec \\ 
  \hline
\end{tabular}
\caption{Program execution time for derivation of third-order and fourth-order correlation matrices of system 2.}\label{tab5}
\setlength{\extrarowheight}{10pt}
\end{table}

Thus, a program has been created that implements the machine derivation of quantum-kinetic equations for describing multiparticle systems. The possibility of using OOP for reproducing analytical schemes for deriving high-order equations of BBGKY quantum chains has been demonstrated. The problem of reducing the time for constructing a quantum-kinetic model of interacting ensembles of different kinds particles has been solved. In particular, the developed OOP algorithms have been applied to the problem of interaction of sets of quantum emitters with continua of quantized electromagnetic field modes.

\section*{Acknowledgement}
The research was carried out within the state assignment of The Ministry of Education of The Russian Federation "Physics of nanostructured materials and highly sensitive sensorics: synthesis, fundamental research and applications in photonics, life sciences, quantum and nanotechnology" (theme No.  - 124031100005-5)

\bibliography{bibliog} 
\end{document}